\documentclass[aps,twocolumn,nofootinbib,showpacs, prd]{revtex4}
\usepackage{graphicx}
\usepackage{amsfonts}
\usepackage{amssymb}
\usepackage{amsbsy}
\usepackage{amsmath}
\usepackage{latexsym}
\usepackage{natbib}
\usepackage{bm}
\usepackage{color}
\usepackage{float}

\begin{document}
\title{Self-similar scalar field collapse}

\author{Narayan Banerjee}
\email{narayan@iiserkol.ac.in}

\author{Soumya Chakrabarti}
\email{adhpagla@iiserkol.ac.in}

\affiliation{ Department of Physical Sciences, 
Indian Institute of Science Education and Research Kolkata,
 Mohanpur Campus, Nadia, West Bengal 741246, India.}
 
\pacs{04.50.Kd; 04.70.Bw}

\date{\today}

\begin{abstract}
A spherically symmetric collapsing scalar field model is discussed with a dissipative fluid which includes a heat flux. This vastly general matter distribution is analysed at the expense of a high degree of symmetry in the spacetime, that of conformal flatness and self similarity. Indeed collapsing models, terminating into a curvature singularity can be obtained. The formation of black holes or the occurrence of naked singularities depend on the initial collapsing profiles.
\end{abstract}

\maketitle

\section{Introduction}
The problem of gravitational collapse of massive stars received a substantial amount of attention for many years, ever since the first significant work in this connection by Oppenheimer and Snyder \cite{Opp}. A spacetime singularity is expected to form as an end-state of a continual gravitational collapse. Though the general speculation is that an event horizon would form and shield this singular state from a distant observer, there are examples with quite reasonable physical conditions, where no horizon can form and the singularity remains exposed, maybe forever, giving rise to what is known as a naked singularity. Therefore it is of great importance to investigate what an unhindered gravitational collapse can lead to. For a systematic exposition of various aspects of gravitational collapse and its possible outcome, we refer to the discussions by Joshi \cite{pankaj1, pankaj2}.                  \\

A recent upsurge of interest in scalar field cosmology is driven by the observational evidence of accelerated expansion of the universe and thus the need for an exotic matter component, called dark energy, which can provide an effective negative pressure \cite{paddy, varun, sami}. A scalar field with a positive potential can indeed serve this purpose. The distribution of the dark energy vis-a-vis the distribution of the fluid remains unclear except perhaps the dogma that it does not cluster at any scale below the Hubble scale. An exhaustive investigation regarding a scalar field collapse might lead to some hint towards the possible clustering of the dark energy. For instance, the spherical collapse model was studied by Mota and Bruck \cite{mota} in the context of dark energy cosmologies, in which dark energy was modelled as a minimally coupled scalar field for different popular interaction potentials. Moreover, formally a scalar field works well as a description of many perfectly reasonable matter distributions as discussed by Goncalves and Moss \cite{gong2}. This serves as an additional motivation for the investigation regarding various aspects of a scalar field distribution.     \\

Some studies are however, already there in the literature in connection with exact solutions of Einstein's equations when a scalar field is coupled to gravity and collapsing modes of such solutions. Christodoulou established the global existence and uniqueness of the solutions of Einstein-scalar field equations \cite{christo1}. A sufficient condition for the formation of a trapped surface in the evolution of a given initial data set was also studied by the same author \cite{christo3}. Goldwirth and Piran showed that a scalar field collapse leads to a singularity which is cut-off from the exterior observer by an event horizon \cite{piran}. Scalar field collapse have been analytically studied by Goswami and Joshi \cite{goswami}, Giambo \cite{giambo} quite recently. Various massive scalar field models exist as a possible driver of the recent accelerated expansion of the universe. A few examples of collapse scenario of such type of scalar fields are already in literature \cite{giambo, gong1, gong2, goswami2, koyel, cai1, cai2}.   \\

Numerical investigations have also provided useful insights into black hole formation in a scalar field collapse. In their extensive works, Choptuik, Brady, and Gundlach \cite{chop, brady, gund} considered the numerical evolution of collapsing profiles characterised by a single parameter $(p)$ and showed that the behaviour of the resulting families of solutions depends on the value of p; there is a critical evolution with $p = p^{\ast}$, which signals the transition between complete dispersal and black hole formation. For a comprehensive review of the critical phenomena associated with a scalar field collapse, we refer to the work of Gundlach\cite{Gundlach1}.  \\

The aim of the present work is to look at the collapse of a massive scalar field along with a fluid distribution which is locally anisotropic and contains a radial heat flux. The potential is taken to be a power law function of the scalar field or suitable combinations of power-law terms. Power law potentials are quite relevant and well-studied in a cosmological context. Many realistic matter distribution can be modelled with power law potentials. For instance, a quadratic potential, on the average, mimics a pressureless dust whereas the quartic potential exhibits radiation like behavior \cite{samibook}. For a potential with a power less than unity results in an inverse power-law self interaction as in the wave equation, the $\frac{dV}{d\phi}$ term introduces a term with inverse power of $\phi$. Inverse power law potentials were extensively used for the construction of tracker fields\cite{zlatev, johri}. We study the scalar field evolution equation extensively for some reasonable choices of potentials so that we have examples from both positive and negative self interactions.           \\

Analytical studies are limited to very special cases due to the nonlinearity of the system of equations. We start with a conformally flat spacetime. In a very recent work, an exact solution for such a scalar field collapse with a spatial homogeneity was given \cite{scnb}, where it was shown that a massive scalar field collapse without any apriori choice of an equation of state, can indeed lead to the formation of a black hole. Under the assumption that the scalar field evolution equation is integrable, the work extensively utilised the integrability conditions for a general anharmonic oscillator, developed by Euler \cite{euler} and utilized by Harko, Lobo and Mak\cite{harko}. For the sake of simplicity, both the conformal factor and the scalar field were assumed to be spatially homogeneous, making the system effectively a scalar field analogue of the Oppenheimer-Snyder collapsing model. In the present work, the assumption of homogeneity is dropped and also the fluid content has a local anisotropy and also has a radial heat flux. So the system described in the present work is a lot more general than the one in \cite{scnb}. However, the price paid is that the present work assumes the existence of a homothetic Killing vector implying a self-similarity in the spacetime.            \\

Indeed the assumption of self-similarity imposes a restriction on the metric tensor, but it is not really unphysical and there are lots of examples where self-similarity is indeed observed. In non-relativistic Newtonian fluid dynamics, self-similarity indicates that the physical variables are functions of a dimensioness variable $\frac{x}{l(t)}$ where $x$ and $t$ are space and time variables and $l$, a function of $t$, has the dimension of length. Existence of self-similarity indicates that the spatial distribution of physical variables remains similar to itself at all time. Such examples can be found in strong explosion and thermal waves. In general relativity also, self-similarity, characterized by the existence of a homethetic Killing vector, finds application. We refer to the work of Carr and Coley\cite{carr} for a review on the implications of self-similarity in general relativity.        \\

The paper is organized as follows. In the second section we write down the relevant equations for a general scalar field model along with a fluid distribution in a conformally flat spherically symmetric spacetime. The third section introduces the self-similarity in the system where we write the relevant equation in terms of a single variable. In the fourth section, the theorem of integrability of an anharmonic oscillator equation is reviewed in brief. We investigate the collapse of a scalar field with a power-law potential in detail in section five. The discussion on a combination of a quadratic and an arbitrary power-law potential is given in section six. The seventh and the final section includes a summary of the results obtained.

\section{Conformally flat Scalar field model with pressure anisotropy and heat flux}
The space-time metric is chosen to have vanishing Weyl tensor implying a conformal flatness. Conformally flat space-times are well-studied and utilised in the context of radiating fluid spheres or shear-free radiating stars \cite{som, maiti, modak, bhui, patel, schafer, ivanov, herrera}. The metric can be written as  \\
\begin{equation}
\label{metric}
ds^2=\frac{1}{{A(r,t)}^2}\Bigg[dt^2-dr^2-r^2d\Omega^2\Bigg],
\end{equation}
where $A(r,t)$ is the conformal factor and governs the evolution of the sphere.        \\

The fluid inside the spherically symmetric body is assumed to be locally anisotropic along with the presence of heat flux. Thus the energy-momentum tensor is given by
\begin{equation}\label{EMT}
T_{\alpha\beta}=(\rho+p_{t})u_{\alpha}u_{\beta}-p_{t}g_{\alpha\beta}+ (p_r-p_{t})\chi_{\alpha} \chi_{\beta}+q_{\alpha}u_{\beta}+q_{\beta}u_{\alpha},
\end{equation}
where $q^{\alpha}=(0,q,0,0)$ is the radially directed heat flux vector. $\rho$ is the energy density, $p_{t}$ the tangential pressure, $p_r$ the radial pressure, $u_{\alpha}$ the four-velocity of the fluid and $\chi_{\alpha}$ is the unit four-vector along the radial direction. The vectors $u_{\alpha}$ and $\chi_{\alpha}$ are normalised as
\begin{equation}
\label{norm}
u^{\alpha}u_{\alpha}=1,\quad\chi^{\alpha}\chi_{\alpha}=-1,\quad\chi^{\alpha}u_{\alpha}=0.
\end{equation}

A comoving observer is chosen, so that $u^{\alpha} = A{\delta}^{\alpha}_{0}$ and the normalization equation (\ref{norm}) is satisfied ($A = \frac{1}{\sqrt{g_{00}}}$). It is to be noted that there is no apriori assumption of an isotropic fluid pressure. The radial and transverse pressures are different. Anisotropic fluid pressure is quite relevant in the study of compact objects and considerable attention has been given to this in existing literature. Comprehensive review can be found in the work of Herrera and Santos\cite{santos1} and Herrera et. al.\cite{herre2}. In addition to the obvious relevance of an anisotropic pressure in collapsing bodies or in any compact object, it has also been shown by Herrera and Leon that on allowing a one-parameter group of conformal motions, a smooth matching of interior and exterior geometry is possible iff there is an pressure anisotropy in the fluid description\cite{herre3}.      \\

We also include a dissipative process, namely heat conduction in the system. In the evolution of stellar bodies, dissipative processes are of utmost importance. Particularly when a collapsing star becomes too compact, the size of the constituent particles can no longer be neglected in comparison with the mean free path, and dissipative processes can indeed play a vital role, in shedding off energy so as to settle down to a stable system. For more relevant details we refer to the works \cite{herre2, kaza}.

\par When a scalar field $\phi=\phi(r,t)$ is minimally coupled to gravity, the relevant action is given by 
\begin{equation}
\textit{A}=\int{\sqrt{-g}d^4x[R+\frac{1}{2}\phi^\mu\phi_\mu-V(\phi)+L_{m}]},
\end{equation}

where $V(\phi)$ is the potential and $L_{m}$ is the Lagrangian density for the fluid distribution.

\par From this action, the contribution to the energy-momentum tensor from the scalar field $\phi$ can be  written as
\begin{equation}
T^\phi_{\mu\nu}=\partial_\mu\phi\partial_\nu\phi-g_{\mu\nu}\Bigg[\frac{1}{2}g^{\alpha\beta}\partial_\alpha\phi\partial_\beta\phi-V(\phi)\Bigg]. 
\end{equation}

Einstein field equations (in the units $8 \pi G = 1$) can thus be written as
\begin{eqnarray}\nonumber
\label{fe1}
&&3\dot{A}^2-3A'^2+2AA''+\frac{4}{r}AA'=\rho+\frac{1}{2}A^2\dot{\phi}^2-\frac{1}{2}A^2\phi'^2 \\&&
+ V(\phi),
\end{eqnarray}

\begin{eqnarray}\nonumber
\label{fe2}
&&2\ddot{A}A-3\dot{A}^2+3A'^2-\frac{4}{r}AA'=p_{r}+\frac{1}{2}{\phi'}^2A^2+\frac{1}{2}A^2\dot{\phi}^2 \\&&
-V(\phi),
\end{eqnarray}

\begin{eqnarray}\nonumber
\label{fe3}
&&2\ddot{A}A-3\dot{A}^2+3A'^2-\frac{2}{r}AA'-2AA''=p_{t}-\frac{1}{2}{\phi'}^2A^2 \\&&
+\frac{1}{2}A^2\dot{\phi}^2-V(\phi),
\end{eqnarray}

\begin{equation}
\label{fe4}
\frac{2\dot{A}'}{A}=-\frac{q}{A^3}+\dot{\phi}\phi'.
\end{equation}

\par The wave equation for the scalar field is given by
\begin{equation}
\label{wave}
\Box\phi+\frac{dV}{d\phi}=0,
\end{equation}
which, for the present metric (\ref{metric}), translates into
\begin{equation}
\label{wave2}
\ddot{\phi}-\phi''-2\frac{\dot{A}}{A}\dot{\phi}-2\frac{\phi'}{r}+2\frac{\phi'A'}{A}+\frac{1}{A^2}\frac{dV}{d\phi}=0.
\end{equation}
$\rho$, $p_{r}$, $p_{t}$, $q$, $A$ and $\phi$ are unknowns in this system of equations containing four equations $(6-8)$. The equation $(9)$ follows from the field equations as a consequence of the Bianchi identities and therefore is not independent. Rather than assuming any specific equation of state to close the system of equation, we probe the system under the assumption that the scalar field equation (\ref{wave2}) is integrable; which facilitates the appearence of an additional differential condition on the conformal factor, as discussed in the following sections.

\section{Self Similarity and exact solution}
A self-similar solution is one in which the spacetime admits a homothetic killing vector $\xi$, which satisfies the equation
\begin{equation}
L_{\xi}g_{ab}=\xi_{a;b}+\xi_{b;a}=2g_{ab},
\end{equation}
where $L$ denotes the Lie derivative. In such a case, one can have repeatative structures at various scales. With a conformal symmetry the angle between two curves remains the same and the distance between two points are scaled depending on the spacetime dependence of the conformal factor ($\frac{1}{A}$ in the present case). For a self-similar space-time, by a suitable transformation of coordinates, all metric coefficients and dependent variables can be put in the form in which they are functions of a single independent variable, which is a dimensionless combination of space and time coordinates; for instance, in a spherically symmetric space this variable is $\frac{t}{r}$.      \\

It deserves mention that a homothetic Killing vector (HKV) is a special case of a conformal Killing vector (CKV) $\eta$ defined by $L_{\eta}g_{ab}=\eta_{a;b}+\eta_{b;a}=\lambda g_{ab}$, where $\lambda$ is a function. Clearly HKV is obtained from a CKV when $\lambda =2$. For a brief but useful discussion on CKV, we refer to the work by Maartens and Maharaj\cite{maartens}.

Writing $A(r,t)=rB(z)$, the scalar field equation (\ref{wave2}) is transformed into

\begin{equation}\label{wave3}
\phi^{\circ\circ}-\phi^{\circ}\Bigg[2\frac{B^{\circ}}{B}+\frac{2z}{1-z^2}\Bigg]+\frac{\frac{dV}{d\phi}}{B^2(1-z^2)}=0.
\end{equation}           \\

Here, an overhead $\circ$ denotes a derivative with respect to $z=\frac{t}{r}$.                      \\

Equation (\ref{wave3}) can be formally written in a general classical anharmonic oscillator equation with variable coefficients as 
\begin{equation}
\label{gen}
\phi^{\circ\circ}+f_1(z)\phi^{\circ}+ f_2(z)\phi+f_3(z)\frac{dV}{d\phi}=0,
\end{equation}

where $f_i$'s are functions of $z$ only.

\section{A note on the integrability of an anharmonic oscillator equation}
A nonlinear anharmonic oscillator with variable coefficients and a power law potential can be written in a general form as

\begin{equation}
\label{gen}
\ddot{\phi}+f_1(u)\dot{\phi}+ f_2(u)\phi+f_3(u)\phi^n=0,
\end{equation}

where $f_i$'s are functions of $u$ and $n \in {\cal Q}$, is a constant. Here $u$ is the independent variable used only to define the theorem. An overhead dot represents a differentiation with respect to $u$. Using Euler’s theorem on the integrability of the general anharmonic oscillator equation \cite{euler} and recent results given  by Harko {\it et al} \cite{harko}, this equation can be integrated under certain conditions. The essence is given in the form of a theorem as \cite{euler, harko},

\textbf{Theorem} If and only if $n\notin \left\{-3,-1,0,1\right\} $, and the coefficients of Eq.(\ref{gen}) satisfy the differential condition
\begin{eqnarray}\nonumber \nonumber
\label{int-gen}
&&\frac{1}{(n+3)}\frac{1}{f_{3}(u)}\frac{d^{2}f_{3}}{du^{2}%
}-\frac{n+4}{\left( n+3\right) ^{2}}\left[ \frac{1}{f_{3}(u)}\frac{df_{3}}{du%
}\right] ^{2} \\&& \nonumber
+ \frac{n-1}{\left( n+3\right) ^{2}}\left[ \frac{1}{f_{3}(u)}%
\frac{df_{3}}{du}\right] f_{1}\left( u\right) + \frac{2}{n+3}\frac{df_{1}}{du} \\&&
+\frac{2\left( n+1\right) }{\left( n+3\right) ^{2}}f_{1}^{2}\left( u\right)=f_2(u),
\end{eqnarray} 
equation (\ref{gen}) is integrable. \\

If we introduce a pair of new variables $\Phi$ and $U$ given by 
\begin{eqnarray}
\label{Phi}
\Phi\left( U\right) &=&C\phi\left( u\right) f_{3}^{\frac{1}{n+3}}\left( u\right)
e^{\frac{2}{n+3}\int^{u}f_{1}\left( x \right) dx },\\
\label{U}
U\left( \phi,u\right) &=&C^{\frac{1-n}{2}}\int^{u}f_{3}^{\frac{2}{n+3}}\left(
\xi \right) e^{\left( \frac{1-n}{n+3}\right) \int^{\xi }f_{1}\left( x
\right) dx }d\xi ,
\end{eqnarray}%

where $C$ is a constant, equation (\ref{gen}) can be written as 

\begin{equation}
\label{Phi1}
\frac{d^{2}\Phi}{dU^{2}}+\Phi^{n}\left( U\right) =0.
\end{equation}

\section{Power Law Potential : $V(\phi)=\frac{\phi^{m+1}}{(m+1)}$}
For a simple power law potential, for which $\frac{dV}{d\phi}=\phi^m$, one can write equation (\ref{wave3}) as
\begin{equation}\label{wave4}
\phi^{\circ\circ}-\phi^{\circ}\Bigg[2\frac{B^{\circ}}{B}+\frac{2z}{1-z^2}\Bigg]+\frac{\phi^m}{B^2(1-z^2)}=0.
\end{equation}           \\
Comparing with (\ref{gen}), it is straightforward to identify $f_{1}(z)=-\Bigg[2\frac{B^{\circ}}{B}+\frac{2z}{1-z^2}\Bigg]$, $f_{2}(z)=0$ and $f_{3}(z)=\frac{1}{B^2(1-z^2)}$. The integrability criteria as mentioned in the last section, gives a second order non-linear differential condition on $B(z)$ as
\begin{eqnarray}\nonumber \label{solbasic}
&&\frac{B^{\circ\circ}}{B}-3\frac{(m+1)}{(m+3)}\Bigg(\frac{B^{\circ}}{B}\Bigg)^2-\frac{8(2m+3)}{6(m+3)}\frac{B^{\circ}z}{B(1-z^2)}\\&&
-\frac{2(1+m)z^2-2(m+3)}{6(m+3)(1-z^2)^2}=0.
\end{eqnarray}
Here $[m\neq{-3,-1,0,1}]$ as mentioned before.

\begin{figure}[h]\label{1stfig}
\begin{center}
\includegraphics[width=0.42\textwidth]{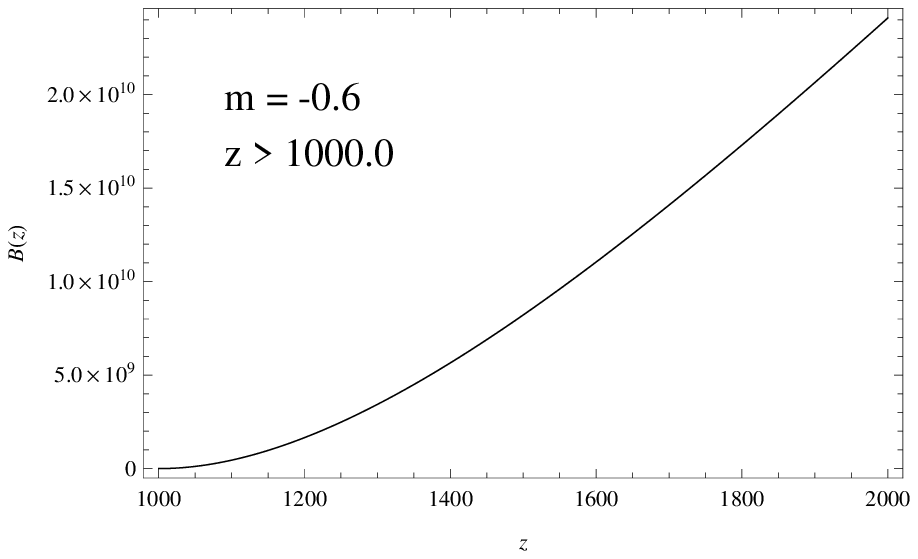}
\includegraphics[width=0.40\textwidth]{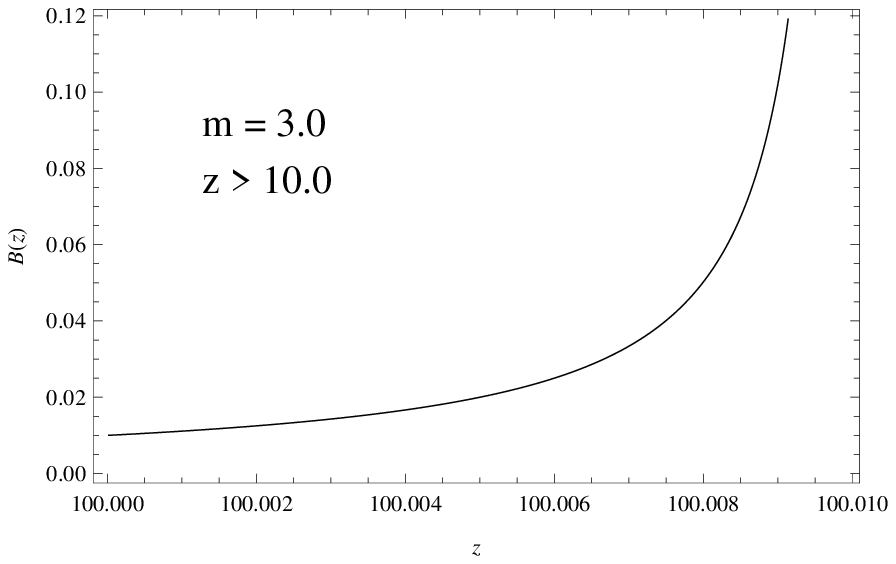}
\caption{Evolution of the conformal factor $B(z)$ with respect to $z$ for $z > 10$. The potentials are taken to be $V(\phi)= \frac{5\phi^{2/5}}{2}$ (graph on the left side) and $V(\phi)= \frac{\phi^4}{4}$ (graph on the right side).}
\end{center}
\end{figure}

It is indeed difficult to handle the equation (\ref{solbasic}) analytically, but it can be treated numerically so as to examine the nature of $B$ as a function of $z$. Our aim is to understand whether the system reaches any singularity of zero proper volume at any finite future.    \\

Figure $(1)$ shows the numerical evolution of $B(z)$ with respect to $z$ for $m=-\frac{3}{5}$ and $m=3$. Keeping in mind the fact that the scale factor is proportional to $\frac{1}{B(z)}$; it appears that for $V(\phi)= \frac{5\phi^{2/5}}{2}$, the system undergoes a uniform gravitational collapse, but the rate of collapse dies down eventually, and a zero proper volume singularity is reached only for $z \rightarrow \infty$, i.e. for finite $t = t_s$ but $r \rightarrow 0$. However, for $V(\phi)= \frac{\phi^4}{4}$, the system shrinks to zero rather rapidly, where $B(z) \rightarrow \infty$, at a finite value of $z$ and this singularity is not necessarily a central singularity.    \\

These plots, though represent the general evolution of the spherical body, further analysis like formation of an apparent horizon or the nature of curvature scalar may not be quite straightforward, without any solution for $B(z)$ in a closed form. With this in mind, we now investigate whether the collapsing system reaches any central singularity at a future time given by $t \rightarrow t_{s}$ and $r \rightarrow 0$. For that purpose we now look for an approximate but analytical expression for $B(z)$ for $z >> 1$. In this domain the last term on the LHS of (\ref{solbasic}), $\frac{2(1+m)z^2-2(m+3)}{(m+3)^2 (1-z^2)^2}$ is of the order of $\sim \frac{1}{z^2}+\frac{1}{z^4}$ and hence can be ignored with respect to the other terms. This approximation does not affect the nature of the evolution as shall be seen in the subsequent analysis. Thus we write the effective equation governing the collapsing fluid for $z >> 1$ as
 
\begin{equation}\label{sol2}
\frac{B^{\circ\circ}}{B}-3\frac{(m+1)}{(m+3)}\Bigg(\frac{B^{\circ}}{B}\Bigg)^2-\frac{8(2m+3)}{6(m+3)}\frac{B^{\circ}z}{B(1-z^2)}=0.
\end{equation}

A solution of equation (\ref{sol2}) can be written in term of Gauss' Hypergeometric function as
\begin{equation}\label{hyper}
\frac{B^{1-\alpha}}{(1-\alpha)}= {_2}F{_1}\Bigg[\frac{1}{2},\beta;\frac{3}{2};z^2\Bigg]\varepsilon z + \varepsilon_0, 
\end{equation}

where $\alpha$ and $\beta$ are defined in terms of $m$ as $\alpha=3\frac{(m+1)}{(m+3)}$, $\beta=\frac{8(2m+3)}{6(m+3)}$ and  $\varepsilon$ and $\varepsilon_0$ are constants of integration. (\ref{hyper}) describes the evolution of the spherical body as a function of the self-similarity variable $z$ and the exponent of the self interaction, $m$. One can expand equation (\ref{hyper}) in a power series in the limit $z \rightarrow \infty$ and write $B(z)$ explicitly as a function of $z$. However, in order to present some simple examples we choose three values of $m$ such that the parameter $\beta$ has the values $1$, $2$ and $0$ respectively; namely, $m=-\frac{3}{5}$, $m=3$ and $m= -\frac{3}{2}$. Therefore the potentials are effectively chosen as $V(\phi)= \frac{5\phi^{2/5}}{2}$, $V(\phi)= \frac{\phi^4}{4}$ and $V(\phi)= -\frac{2}{\phi^{1/2}}$. We note here that a case where $m = -\frac{3}{2}$ is also taken up so as to have an example for an inverse power law potential as well.

\begin{enumerate}
\item{{\bf Case 1} : {\bf $m=-\frac{3}{5}$, $\beta = 1$ and $\alpha = \frac{1}{2}$} \\

Putting these values in (\ref{sol2}), a solution can be written as
\begin{equation}\label{exact-sol}
B(z)=C_{2}\Bigg[C_{1}+\frac{1}{2} ln\Big|z+(z^2-1)^\frac{1}{2}\Big|\Bigg]^2,
\end{equation}            
where $C_{1}$ and $C_{2}$ are constants of integration. Therefore the inverse of the conformal factor can be written explicitly as a function of $r$ and $t$ as
\begin{equation}\label{exact-sol2}
A(z)=rC_{2}\Bigg[C_{1}+\frac{1}{2} ln\Bigg|\frac{t}{r}+\Bigg(\frac{t^2}{r^2}-1\Bigg)^\frac{1}{2}\Bigg|\Bigg]^2.
\end{equation} 

\begin{figure}[h]\label{2ndfig}
\begin{center}
\includegraphics[width=0.42\textwidth]{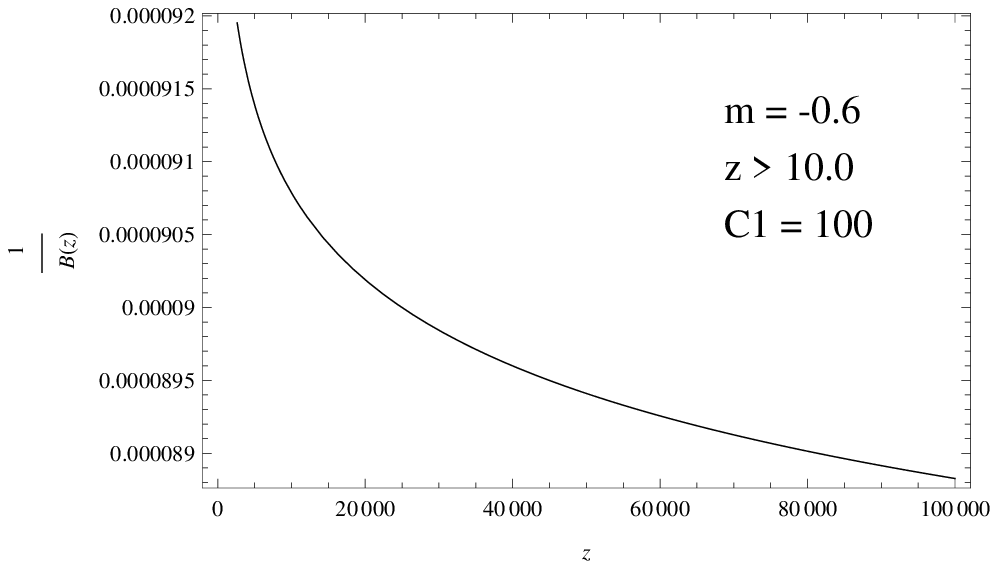}
\includegraphics[width=0.40\textwidth]{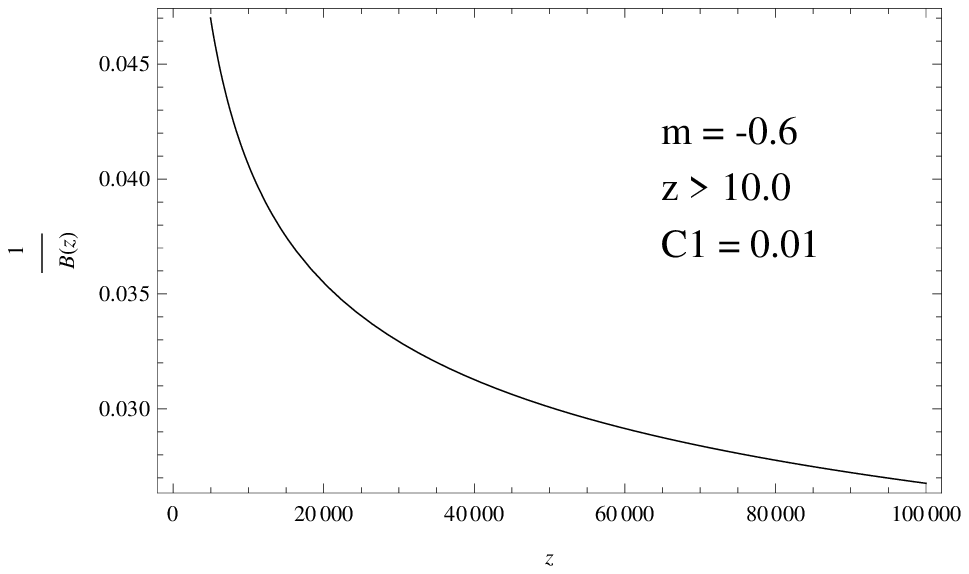}
\includegraphics[width=0.40\textwidth]{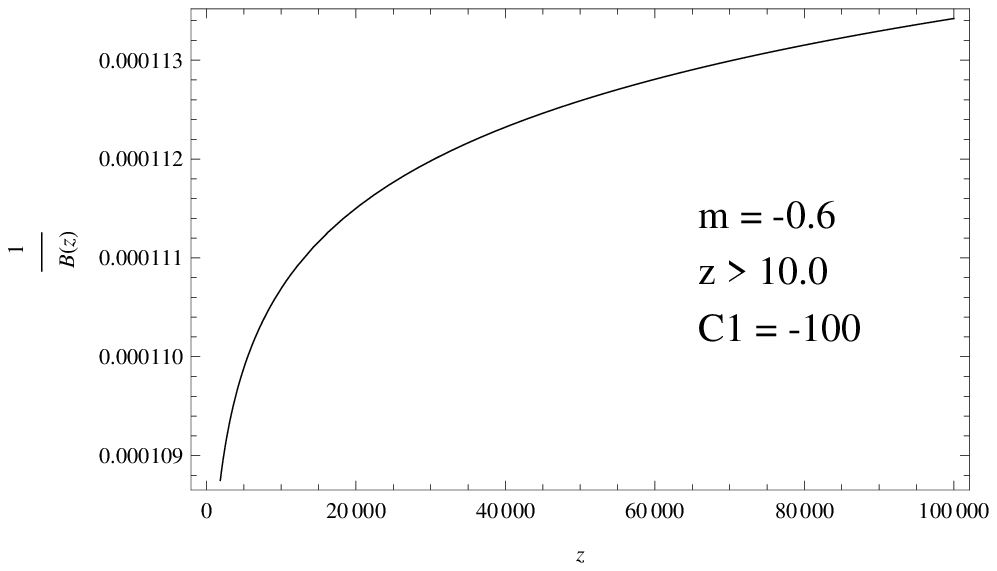}
\caption{Evolution of $B(z)$ with respect to $z$ for $V(\phi)= \frac{5\phi^{2/5}}{2}$, for different choices of the parameter $C_1$.}
\end{center}
\end{figure}

The evolution with respect to $z$ and therefore formation of a singularity will depend on the integration constants and of course, the choice of potential does play a crucial role here. Without any loss of generality we assume $C_{2}=1$. It must be mentioned, that for all $C_{2}<0$, one has either a negative volume or no real evolution at all. Excluding those cases, now we examine equation (\ref{exact-sol}) near the central singularity. It is evident that, the singularity, characterised by $B \rightarrow \infty$, appears only when $z \rightarrow \infty$, i.e, either when $t_s \rightarrow \infty$, or when $t \rightarrow t_s$, $r \rightarrow 0$. However, for all negative values of $C_1$, the system in not collapsing at all; rather, it settles down asymptotically at a finite volume after a period of steady expansion with respect to $z$. Figure $(2)$ shows the evolution of $B(z)$ with respect to $z$.
}
\item{{\bf Case2: } {\bf $m=3$, $\alpha = \beta = 2$}. \\

With these values of the parameters, equation (\ref{sol2}) can be solved to write $B(z)$ as
\begin{equation}\label{exact-sol3}
B(z)= \frac{C_2}{ln{\Big[\frac{C_1(1+z)}{(1-z)}\Big]}}.
\end{equation}
Here, $C_2$ can again be chosen to be unity. For the solution to be valid in the region $z >> 1$, $C_1$ must be a non-zero negative number such that $ln{\Big[\frac{C_1(1+z)}{(1-z)}\Big]}$ is real.

\begin{figure}[h]\label{3rdfig}
\begin{center}
\includegraphics[width=0.40\textwidth]{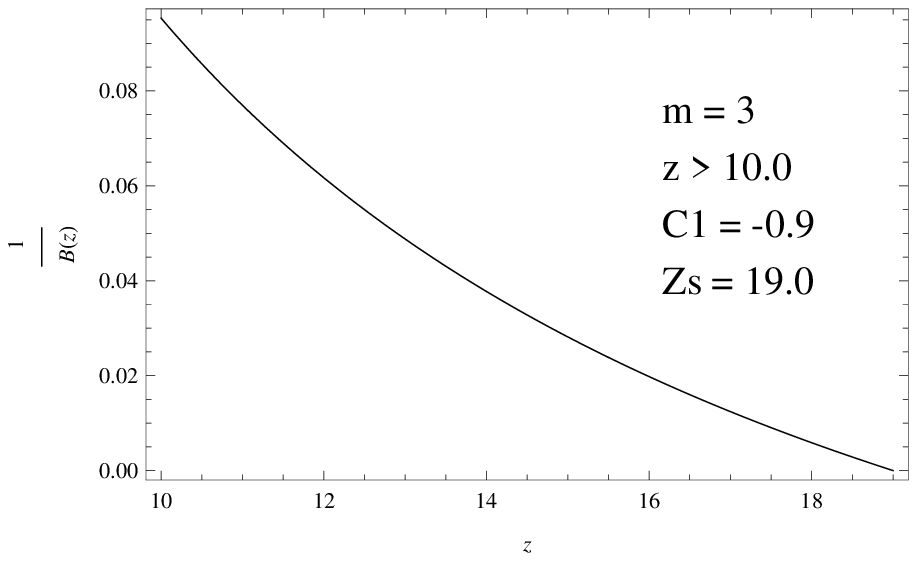}
\includegraphics[width=0.42\textwidth]{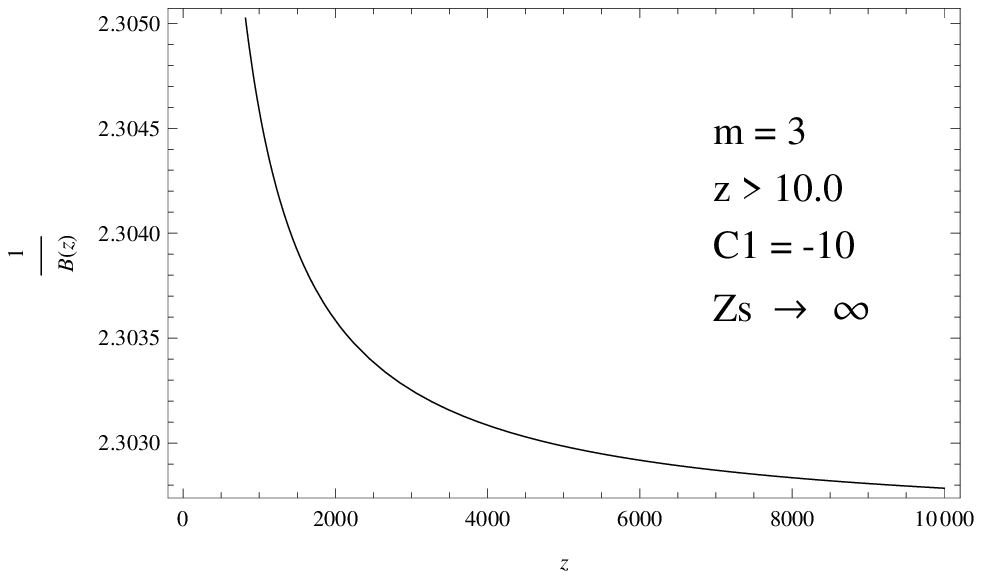}
\caption{Evolution of $B(z)$ with respect to $z$ for $V(\phi)= \frac{\phi^4}{4}$ for different negative values of the parameter $C_1$.}
\end{center}
\end{figure}

From (\ref{exact-sol3}) one can find that the system reaches zero proper volume singularity at a finite value of $z$ given by
\begin{equation}
z_s= \frac{1-C_1}{1+C_1}.
\end{equation}
However, for a collapsing model, the scale factor must be a decreasing function throughout. This means $\frac{d}{dz} ln\Big[\frac{(1+z)C_1}{(1-z)}\Big] < 0$. On simplification, this yields $\frac{2}{(1-z^2)} < 0$, which holds true iff $z > 1$. So $z_s$, defined by $z_s = \frac{1-C_1}{1+C_1}$ is greater than $1$, only if $-1 < C_1 < 0$, which is, therefore consistent with thr requirement of a negative ${C_{1}}$. The discussion is supported graphically in figure $(3)$.
}

\item{{\bf Case $3$ : $m= -\frac{3}{2}$} {\bf $\alpha= -1$ and $\beta= 0$.} \\

Equation (\ref{sol2}) is simplified significantly and solution of $B(z)$ may be written as
\begin{equation}\label{exact-sol4}
B(z)= [2 C_1 (z+z_0)]^{\frac{1}{2}}.
\end{equation} 
Here, $C_1 > 0$ and $z_0$ are constants of integration. The scale factor is defined as $Y(r,t) = \frac{1}{[2 C_1 (\frac{t}{r}+z_0)]^{\frac{1}{2}}}$. 
\begin{figure}[h]\label{3rdfig}
\begin{center}
\includegraphics[width=0.40\textwidth]{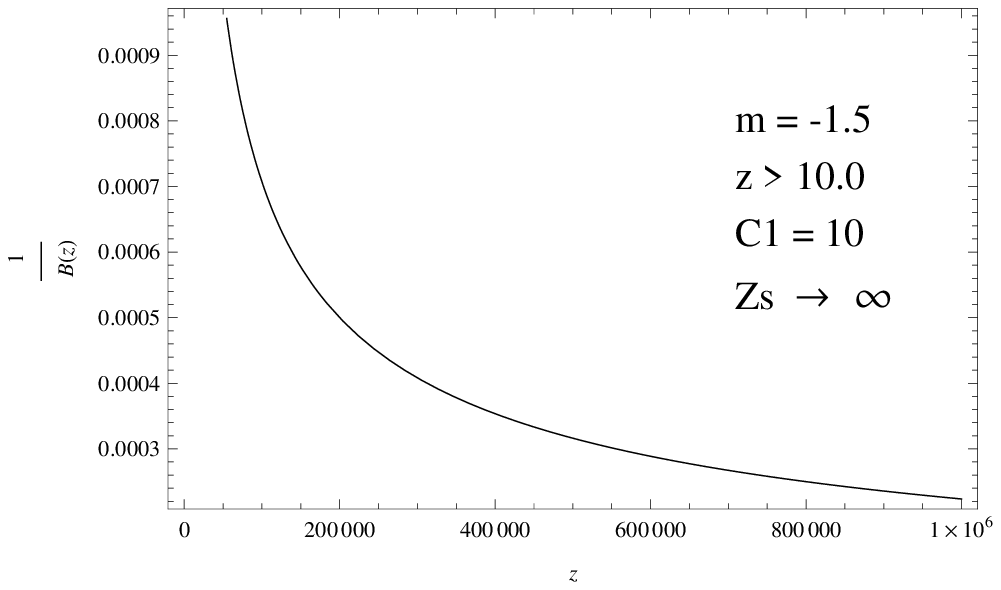}
\includegraphics[width=0.42\textwidth]{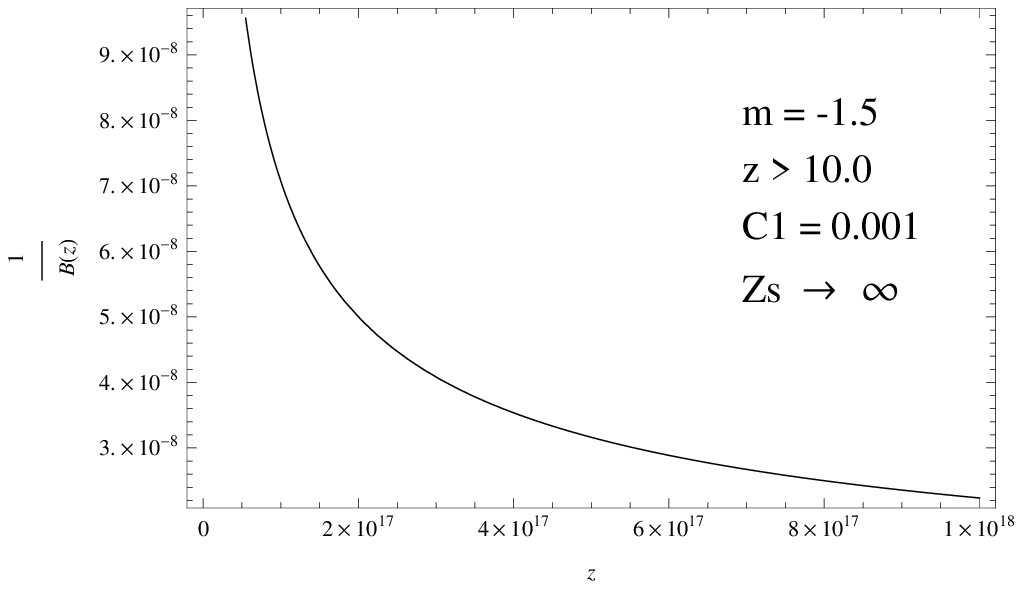}
\caption{Evolution of $B(z)$ with respect to $z$ for $V(\phi)= -\frac{2}{\phi^{1/2}}$ for different positive values of $C_1$. The qualitative behaviour remains the same over different choices of the parameter.}
\end{center}
\end{figure}
Now, $Y \rightarrow 0$ only when $z \rightarrow \infty$, i.e., $t \rightarrow t_s$, $r \rightarrow 0$. Thus the system reaches a central singularity ar $t \rightarrow t_s$, as shown in figure $(4)$ for different choices of initial conditions defined by the choice of $C_1$.         
}
\end{enumerate}

\subsection{Expressions for scalar field, physical quantities and curvature scalar}
Point transforming the scalar field equation using (\ref{Phi}) and (\ref{U}), equation (\ref{wave3}) can be written in an integrable form as (\ref{Phi1}). Using the exact form of the coefficients $f_{1}(z)$ and $f_{3}(z)$, after some algebra, we express the scalar field as a function of $z$ as
\begin{eqnarray}\nonumber \label{scalarfield}
&&\phi(z)=\phi_{0}B^{\frac{6}{(m+3)}}(1-z^2)^{-\frac{1}{(m+3)}}\Bigg[C^{\frac{(1-m)}{2}} \\&&
{\int B^{\frac{2(m-3)}{(m+3)}}(1-z^2)^{-\frac{(m+1)}{(m+3)}}dz}-\Phi_0\Bigg].
\end{eqnarray}
Here, $C$ is an integration constant coming from the definition of the point transformation and $\phi_{0}$ is defined in terms of $C$. $\Phi_0$ is a constant coming from integration over $z$. For all choices of $m$ discussed in the previous section, $B^{\frac{6}{(m+3)}}$ is an increasing function w.r.t $z$, and diverges when $B(z) \rightarrow \infty$. Thus it is noted that the scalar field diverges at the singularity.       \\

From (\ref{fe1}), (\ref{fe2}), (\ref{fe3}) and (\ref{fe4}), the density, radial and tangential pressure and the heat flux can be expressed generally as
\begin{eqnarray}\nonumber \label{denst}
&&\rho=3\dot{A}^2-3A'^2+2A''A+\frac{4}{r}A'A-\frac{1}{2}A^2\dot{\phi}^2+\frac{1}{2}A^2\phi'^2\\&&
-\frac{\phi^{m+1}}{(m+1)}.
\end{eqnarray}

\begin{eqnarray}\nonumber \label{radpress}
&&p_{r}=2\ddot{A}A-3\dot{A}^2+3A'^2-\frac{4}{r}A'A-\frac{1}{2}\phi'^2A^2-\frac{1}{2}\dot{\phi}^2A^2\\&&
+\frac{\phi^{m+1}}{(m+1)}.
\end{eqnarray}

\begin{eqnarray}\nonumber \label{tangpress}
&&p_{t}=2\ddot{A}A-3\dot{A}^2+3A'^2-2A''A-\frac{2}{r}A'A+\frac{1}{2}\phi'^2 A^2\\&&
-\frac{1}{2}\dot{\phi}^2A^2+\frac{\phi^{m+1}}{(m+1)}.
\end{eqnarray}

\begin{equation}\label{heatf}
q=-2\dot{A}'A^2+\dot{\phi}\phi'A^3.
\end{equation}                               

One can use the approximate solution for $B$ from equations (\ref{exact-sol2}), (\ref{exact-sol3}) and (\ref{exact-sol4}) and use the fact that $A(r,t) = rB(z)$ in order to check the nature of the fluid variables at singularity. It can be found that when the sphere shrinks to zero volume, all these quantities diverge to infinity, confirming the formation of a singularity. To comment on the nature of singularity we write the kretschmann curvature scalar from (\ref{metric}) as 
\begin{eqnarray}\nonumber \nonumber \label{kret}
&&K=-7\dot{A}'^2A^2+4(\dot{A}^2-A'^2 +\frac{AA'}{r}+A''A)^2+ \\&& \nonumber
2(\dot{A}^2-A'^2+A''A -\ddot{A}A)^2+4(\dot{A}^2-A'^2+\\ &&
\frac{AA'}{r}-\ddot{A}A)^2+(\dot{A}^2-A'^2+2\frac{AA'}{r})^2.
\end{eqnarray}
For any given time-evolution, the kretschmann scalar diverges anyway when $r\rightarrow 0$, which indicates that the central shell focussing ends up in a curvature singularity. We note that a singularity may also form in relevant cases where $r \neq 0$. To investigate such a singularity, we write \ref{kret} as a function of $z$ given by
\begin{eqnarray}\nonumber \nonumber \nonumber \nonumber
&&K(z)= -7{B^{\circ}}^2z^2B^2 + 4[{B^{\circ}}^2-(B-B^{\circ}z)^2 + \\&& \nonumber
B(B-B^{\circ}z)-B^{\circ\circ}Bz^2]^2 + 2[{B^{\circ}}^2-(B-B^{\circ}z)^2+ \\ && \nonumber
B^{\circ\circ}B-B^{\circ\circ}Bz^2]^2+ 4[{B^{\circ}}^2-(B-B^{\circ}z)^2 \\&& \nonumber
+B(B-B^{\circ}z)-B^{\circ\circ}B]^2 + [{B^{\circ}}^2-(B-B^{\circ}z)^2\\ &&
 + 2B(B-B^{\circ}z)]^2.
\end{eqnarray}
From this above expression we note that the first term on the RHS of Kretschmann scalar ${B^{\circ}}^2z^2B^2$ diverges anyway when $B(z) \rightarrow \infty$. Thus, the singularity always turns out to be a curvature singularity.

\subsection{Apparent Horizon}
Visibility of the central singularity depends on the formation of an apparent horizon, the surface on which outgoing light rays are just trapped, and cannot escape outward. The condition of such a surface is given by
\begin{equation}\label{appdef}
g^{\mu\nu}R,_{\mu}R,_{\nu}=0,
\end{equation}
where $R(r,t)$ is the proper radius of the collapsing sphere, which is $rB(z)$ in the present work.

Writing the derivative in terms of $z= \frac{t}{r}$, one can express (\ref{appdef}) as
\begin{equation}\label{appselfsimilar}
(1-z^2){B^{\circ}}^2=0.
\end{equation}
We investigate the formation of an apparent horizon in all the relevant cases discussed, for $V(\phi)= \frac{5\phi^{2/5}}{2}$, $V(\phi)= \frac{\phi^4}{4}$ and $V(\phi)= -\frac{2}{\phi^{1/2}}$, with the assumption $C_2 = 1$ and $z >> 1$.

\begin{enumerate}
\item{{\bf Case 1: $m=-\frac{3}{5}$} \\
The equation (\ref{appselfsimilar}) gives the condition
\begin{equation}
z+(z^2-1)^{1/2}=e^{-2(C_1-\delta_0^{1/2})}= e^{\gamma},
\end{equation}
where we have defined $\gamma = -2(C_1-\delta_0^{1/2})$ and $\delta_0$ is a constant of integration. The above equation can be simplified to find the time of formation of apparent horizon as
\begin{equation}
t_{ap} = \frac{re^{-\gamma}}{2}(e^{2\gamma}-1)= rsinh{\gamma}.
\end{equation}
Since $e^{-\gamma}$ is always positive, the structure of the spacetime depends on $(e^{2\gamma}-1)$ and therefore on the parameter $\gamma$. 
}
\item{{\bf Case 2: $m=3$} \\
The equation (\ref{appselfsimilar}) is simplified to yield the condition
\begin{equation}
C_{1}\Big(\frac{1+z}{1-z}\Big) = e^{1/{\Psi_0}}.
\end{equation}
Here $\Psi_0$ is a constant of integration over $z$. One can further simplify to write the time of formation of apparent horizon as
\begin{equation}
t_{ap} = r\Bigg(\frac{e^{\frac{1}{\Psi_0}}-C_1}{e^{\frac{1}{\Psi_0}}+C_1}\Bigg) = r\Gamma_0.
\end{equation}
We note here that considering an apparent horizon is not necessary for those cases where singularity forms only when $t \rightarrow \infty$, i.e. when there is no real singularity at all.
}
\item{{\bf Case 3: $m=-\frac{3}{2}$} \\
 
In a similar manner, for $V(\phi)= -\frac{2}{\phi^{1/2}}$, the condition for an apparent horizon may be written as
\begin{equation}
t_{ap} = r\Big(\frac{{\chi_0}^2}{2\tau}-z_0\Big),
\end{equation}
where $\chi_0$ and $\tau$ are constants of integration and are dependent on suitable choice of initial conditions.
}
\end{enumerate}

In all the cases we find that an apparent horizon is formed at a finite time. As there is no explicit expression for $t_{s}$, (the time of formation of the singularity) the visibility of the singularity cannot be ascertained clearly. But it is quite clear that the singularity can not be visible indefinitely. If $t_{app} > t_{s}$, it can be visible for a finite time only.

\subsection{Matching with an exterior Vaidya Solution}
Generally, in collapsing models, a spherically symmetric interior is matched with a suitable exterior solution; Vaidya metric or a Schwarzschild metric depending on the prevailing conditions \cite{dwivedi}. This requires the continuity of both the metric and the extrinsic curvature on the boundary hypersurface. As the interior has a heat transport defined, the radiating Vaidya solution is chosen as a relevant exterior to be matched with the collapsing sphere. The interior metric is defined as
\begin{equation}
\label{metric2}
ds^2=\frac{1}{{A(r,t)}^2}\Bigg[dt^2-dr^2-r^2d\Omega^2\Bigg],
\end{equation}
and the Vaidya metric is given by
\begin{equation}\label{vaidya}
ds^2=\Bigg[1-\frac{2m(v)}{R}\Bigg]dv^2+2dvdR-R^2d\Omega^2.
\end{equation}
The quantity $m(v)$ represents the Newtonian mass of the gravitating body as measured by an observer at infinity. The metric (\ref{vaidya}) is the unique spherically symmetric solution of the Einstein field equations for radiation in the form of a null fluid.  The necessary conditions for the smooth matching of the interior spacetime to the exterior spacetime was presented by Santos \cite{santos} and also discussed in detail by Chan \cite{chan}, Maharaj and Govender \cite{maharaj} in context of a radiating gravitational collapse. Following their work, The relevant equations matching (\ref{metric2}) with (\ref{vaidya}) can be written as                         
\begin{equation}
\frac{r}{A(r,t)}{=^\Sigma}R,
\end{equation}

\begin{equation}
m(v){=^\Sigma}\frac{r^3}{2A^3}\Bigg(\dot{A}^2-A'^2+\frac{A'A}{r}\Bigg),
\end{equation}
and
\begin{equation}\label{prq}
p_r{=^\Sigma}\frac{q}{A(r,t)},
\end{equation}                        
where $\Sigma$ is the boundary of the collapsing fluid. 
\\
The relation between radial pressure and the heat flux as in equation (\ref{prq}) yields a nonlinear differential condition between the conformal factor and the scalar field to be satisfied on the boundary hypersurface $\Sigma$. In view of equations (\ref{radpress}) and (\ref{heatf}) the condition can be written as
\begin{eqnarray}\nonumber
&&\Bigg(2\frac{\ddot{A}}{A}-2\frac{\dot{A}^2}{A^2}+3\frac{A'^2}{A^2}-4\frac{A'}{Ar}+2\frac{\dot{A}'}{A}\Bigg)-\frac{1}{2}(\dot{\phi}+\phi')^2 \\&&
+\frac{1}{A^2}\frac{\phi^{(m+1)}}{(m+1)} {=^\Sigma} 0.
\end{eqnarray} 

It deserves mention that we cannot obtain any analytical expression from the matching conditions and thus their feedback in the interior solutions could not be discussed. Actually the condition of integrability of the scalar field equation has been utilized to find a solution for the metric but the evolution equation for the scalar field defined by (\ref{scalarfield}) could not be explicitly integrated. Therefore the matching condition could not be solved without the expression of scalar field in a closed form. However, for any interior solution, there is a requirement of the matching, so we just discuss the method and obtain the relevant equations to be solved.

\subsection{Nature of the singularity}
In a spherically symmetric gravitational collapse leading to a central singularity, if the centre gets trapped prior to the formation of singularity, then it is covered and a black hole results. Otherwise, it could be naked, when non-space like future directed trajectories escape from it. Therefore the important point is to determine whether there are any future directed non-spacelike geodesics emerging from the singularity.             \\
From this point of view, general relativistic solutions of self-similar collapse of an adiabatic perfect fluid was discussed by Ori and Piran \cite{ori} where it was argued that a shell-focussing naked singularity may appear if the equation of state is soft enough. Marginally bound self-similar collapsing Tolman spacetimes were examined and the necessary conditions for the formation of a naked shell focussing singularity were discussed by Waugh and Lake \cite{waugh}. For a comprehensive description of the mathematical formulations on occurence of naked singularity in a spherically symmetric gravitational collapse, we refer to the works of Joshi and Dwivedi \cite{jdw1, jdw2}, Dwivedi and Dixit \cite{dixit}. Structure and visibility of central singularity with an arbitrary number of dimensions and with a general type I matter field was discussed by Goswami and Joshi \cite{goswami3}. They showed that the space-time evolution goes to a final state which is either a black hole or a naked singularity, depending on the nature of initial data, and is also subject to validity of the weak energy condition. Following the work of Joshi and Dwivedi \cite{jdw1, jdw2}, a similar discussion was given on the occurrence of naked singularities in the gravitational collapse of an adiabatic perfect fluid in self-similar higher dimensional space–times, by Ghosh and Deshkar \cite{ghosh}. It was shown that strong curvature naked singularities could occur if the weak energy condition holds.              \\

One can write a general spherically symmetric metric as
\begin{equation}
ds^2 = e^{\vartheta}dt^2 - e^{\chi}dr^2 - r^2 S^2 d \Omega^2,
\end{equation}
where $\vartheta$, $\chi$ and $S$ are functions of $z = \frac{t}{r}$. This space-time admits a homothetic killing vector $\xi^{a} = r\frac{\partial}{\partial r} + t\frac{\partial}{\partial t}$. For null geodesics one can write $K^{a}K_{a} = 0$, where $K^{a} = \frac{dx^{a}}{dk}$ are tangent vectors to null geodesics. Since $\xi^{a}$ is a homothetic killing vector, one can write 
\begin{equation}
{\mathcal{L}}_{\xi} g_{ab} = \xi_{a;b} + \xi_{b;a} = 2 g_{ab}
\end{equation}
where $\mathcal{L}$ denotes the Lie derivative. Then it is straightforward to prove that $\frac{d}{dk}(\xi^{a}K_a) = (\xi^{a}K_a)_{;a}K^b = 0$ (for mathematical details we refer to \cite{dixit}). Therefore one can write
\begin{equation}
\xi^{a}K_a = C,
\end{equation}
for null geodesics where $C$ is a constant.
From this algebraic equation and the null condition, one gets the following expressions for $K^t$ and $K^r$ as \cite{ghosh}
\begin{equation}\label{kt1}
K^t = \frac{C \left[z \pm e^{\chi} \Pi\right]}{r\left[ e^{ \chi} - e^{\vartheta} z^2 \right]},
\end{equation}

\begin{equation}\label{kr1}
K^r = \frac{C \left[1 \pm z e^{\vartheta} \Pi\right]}{r\left[ e^{\chi} - e^{\vartheta} z^2 \right]}, 
\end{equation}

where $\Pi =\sqrt{e^{-\chi - \vartheta}} > 0$. Radial null geodesics, by virtue of Eqs. (\ref{kt1}) and (\ref{kr1}), satisfy
\begin{equation}\label{de1}
\frac{dt}{dr} = \frac{z \pm e^{ \chi} \Pi}{1 \pm z e^{ \vartheta}\Pi}.
\end{equation}

The singularity that might have been there is at least locally naked if there exist radial null geodesics emerging from the singularity, and if no such geodesics exist it is a black hole. If the singularity is naked, then there exists a real and positive value of $z_{0}$ as a solution to the algebraic equation
\begin{equation}\label{lm1}
z_{0} = \lim_{t\rightarrow t_{s} \atop r\rightarrow 0} z = \lim_{t\rightarrow t_{s} \atop r\rightarrow 0} \frac{t}{r}=\lim_{t\rightarrow t_{s} \atop r\rightarrow 0} \frac{dt}{dr}.
\end{equation}

Waugh and Lake \cite{waugh2} discussed shell-focussing naked singularities in self-similar spacetimes, considering a general radial homothetic killing trajectory. The lagrangian can be written in terms of $V(z) = (e^{\chi}-{z^2}e^{\vartheta})$, whose value determines the nature of the trajectory, for instance, if $V(z) = 0$, the trajectory is null, and for $V(z) > 0$ the trajectory is space-like. In the null case the trajectory can be shown to be geodesic \cite{waugh2}. If $V(z) = 0$ has no roots given by $z = z_{0}$ then the singularity is not naked. On the other hand the central shell focusing is at-least locally naked if $V(z_0) = 0$ admits one or more positive roots. The values of the roots give the tangents of the escaping geodesics near the singularity. \cite{ghosh, jdw1, jdw2}.                \\

For a conformally flat space-time, the metric is defined as $ds^2=\frac{1}{{A(r,t)}^2}\Big[dt^2-dr^2-r^2d\Omega^2\Big]$, so we have $e^{\chi}=e^{\vartheta}= \frac{1}{A^2}$, $V(z)= \frac{1}{A^2}(1-z^2)$ and $\Pi = A^2$. So the equation for radial null geodesic simplifies considerably into $\frac{dt}{dr} = 1$ (where we have considered only the positive solution as we are looking for a radial outgoing ray). We consider the three cases  for which we have found exact collapsing solutions in section $(5)$.

\begin{enumerate}
\item{{\bf Case 1: $m=-\frac{3}{5}$}  \\

Using the time evolution (\ref{exact-sol2}) and the condition that the central shell focusing is at-least locally naked for $V(z_0) = 0$ one finds 
\begin{equation}\label{zz}
z_0+(z_0^2-1)^\frac{1}{2} = e^{2C_1},
\end{equation}

for the conformally flat metric, from which it is straight-forward to write $z_0 = \frac{1}{2}\Big(e^{2C_1}+e^{-2C_1}\Big)$. The visibility of the singularity depends on the existence of positive roots to Eq. (\ref{zz}), therefore on the positivity of $\Big(e^{2C_1}+e^{-2C_1}\Big)$. Since $z_{0} = \lim_{t\rightarrow t_{s} \; r\rightarrow 0} \frac{dt}{dr}$, the equation for the null geodesic emerging from the singularity may be written as
\begin{equation}\label{ngeo}
t-t_{s}(0) = \frac{1}{2}\Big(e^{2C_1}+e^{-2C_1}\Big) r.
\end{equation}
For all values of $C_{1}$, $\Big(e^{2C_1}+e^{-2C_1}\Big)$ is always greater than zero. Therefore it is always possible to find a radially outward null geodesic emerging from the singularity, indicating a naked singularity.
}
\item{{\bf Case 2: $m=3$} \\

Using (\ref{exact-sol3}) in $V(z_0) = 0$, one obtains 

\begin{equation}
\frac{C_{1}(1+z_0)}{(1-z_0)} = 1.
\end{equation}
For this collapsing model, since $C_1$ must be a negative number, and $z_0$ can be written as $z_0 = \frac{1-C_1}{1+C_1}$. Thus one can indeed have a naked singularity for $-1 < C_{1} < 0$. The singularity in this case is realised at a finite value of $z = \frac{t}{r}$, not at $r \rightarrow 0$ and therefore is not a central singularity. This kind of singularities are generally expected to be covered by a horizon \cite{dadhi}.
}
\item{{\bf Case 3: $m=-\frac{3}{2}$}   \\

Equations (\ref{exact-sol4}) and the condition $V(z_0) = 0$ lead to
\begin{equation}
\frac{1}{4 C_{1} z_0} = 0,
\end{equation}
which can never have any finite positive solution for $z_0$. Therefore the central singularity in this particular case, defined at $\frac{1}{A(z)} \rightarrow 0$ at $r \rightarrow 0$, is always hidden to an exterior observer.
}
\end{enumerate}

\par In a recent work Hamid, Goswami and Maharaj \cite{hamid} showed that a spherically symmetric matter cloud evolving from a regular initial epoch, obeying physically reasonable energy conditions is found to be free of shell crossing singularities. A corollary of their work is that for a continued gravitational collapse of a spherically symmetric perfect fluid obeying the strong energy condition $\rho \geq 0$ and $(\rho + 3p) \geq 0$, the end state of the collapse is necessarily a black hole for a conformally flat spacetime. This idea was also confirmed very recently in a massive scalar field analogue of the collapsing model in a conformally flat spacetime \cite{scnb}. The present work, however, is quite different, as it deals with matter where the strong energy condition is not guaranteed. This can perhaps be related to the contribution of the scalar field to the energy momentum tensor and the dissipation part of the stress-energy tensor in the form of a heat flux. The choices of potential and the initial conditions in the form of the constants of integration can indeed conspire amongst themselves so that the energy conditions are violated which leads to the formation of a naked singularity. A violation of energy condition by massive scalar fields is quite usual and in fact forms the basis of its use as a dark energy. In the present case, as discussed, we find various possibilities, and a black hole is not at all the sole possibility, i.e. the central singularity may not always be covered, at least for some specific choices of the self-interacting potential, for instance $V(\phi)=\frac{5\phi^{2/5}}{2}$ as found out in the present work.

\subsection{Non-existence of shear}

For a general spherically symmetric metric

\begin{equation}
\label{metric-gen}
ds^2 = S^2dt^2 - B^2dr^2 - R^2 d\Omega^2,
\end{equation}

where $S, B, R$ are functions of $r, t$, and a comoving observer is defined so that the velocity vector is $u^{\alpha}=S^{-1}\delta_0^{\alpha}$, the shear tensor components can be easily calculated. The net shear scalar $\sigma$ can be found out as

\begin{equation}
\label{shear}
\sigma=\frac{1}{S}\left(\frac{\dot{B}}{B}-\frac{\dot{R}}{R}\right).
\end{equation}

For the conformally flat metric chosen in the present work, $S = B = \frac{R}{r} = \frac{1}{A(r,t)}$. It is straightforward to see that $\sigma = 0$ in the present case.  In the present case, the existence of anisotropic pressure and dissipative processes might suggest the existence of shear in the spacetime, but the situation is actually shearfree. A shearfree motion is quite common in the discussion of gravitational collapse, and it is not unjustified either. But it should be noted that a shearfree condition, particularly in the presence of anisotropy of the pressure and dissipation, leads to instability. This result has been discussed in detail by Herrera, Prisco and Ospino\cite{herre5}.

\section{Combination of power-law potentials : $V(\phi)=\frac{\phi^2}{2}+\frac{\phi^{(m+1)}}{(m+1)}$}
In section $(5)$ we studied the scalar field equations for a choice of potential for which $\frac{dV}{d\phi}=\phi^m$ exploiting the integrability of the scalar field evolution equation. This approach, though works only for those cases where the scalar field equation is integrable, is useful to make some general comments regarding the collapsing situation. However, the domain of the coordinate transformation restricted us not to choose a few values of $m$ ($m\neq{-3,-1,0,1}$). This excluded any chance of studying cases with a quadratic potential. In this section we assume a form for the potential such that $\frac{dV}{d\phi} = \phi + \phi^m$, i.e. $V(\phi)$ is a combination of two power-law terms, one of them being quadratic in $\phi$. With this choice, the integrability criterion yields a non-linear second order differential equation for the conformal factor $B(z)$ as

\begin{eqnarray}\nonumber \label{evolu2}
&&\frac{B^{\circ\circ}}{B}-3\frac{(m+1)}{(m+3)}\Bigg(\frac{B^{\circ}}{B}\Bigg)^2-\frac{8(2m+3)}{6(m+3)}\frac{B^{\circ}z}{B(1-z^2)} \\&&
-\frac{2(1+m)z^2-2(m+3)}{6(m+3)(1-z^2)^2}+\frac{(m+3)}{6B^2(1-z^2)}=0.               
\end{eqnarray}                                   

It is very difficult to find an exact analytical form of $B(z)$ from (\ref{evolu2}) and thus any further analytical investigations regarding the collapsing geometry is quite restricted. However, we make use of a numerical method to analyse the evolution which of course depends heavily on choice of initial values and ranges of $z$, $B(z)$ and $\frac{dB}{dz}$. First the equation (\ref{evolu2}) is written in terms of $D(z) = \frac{1}{B(z)}$ as

\begin{eqnarray}\nonumber \label{devolu2}
&&\frac{D^{\circ\circ}}{D}= \Bigg[2-\frac{3(m+1)}{m+3}\Bigg]\Bigg(\frac{D^{\circ}}{D}\Bigg)^2+\frac{8(2m+3)}{6(m+3)}\frac{D^{\circ}z}{D(1-z^2)}\\&&
-\frac{2(1+m)z^2-2(m+3)}{6(m+3)(1-z^2)^2}+\frac{(m+3)D^2}{6(1-z^2)}=0.
\end{eqnarray}

Now equation (\ref{devolu2}) is solved numerically and studied graphically as $D(z)$ vs $z$ in the limit $z >>1$. Since we mean to study a collapsing model, $D^{\circ}(z)$ is always chosen as negative. The evolution of the collapsing sphere is sensitive to the factor $\mid\frac{D(z)}{D^{\circ}(z)}\mid$, at least for some choices of potential as will be shown in the subsequent analysis. We present the results obtained in three different categories.       \\  

\begin{figure}[t]\label{RK41}
\begin{center}
\includegraphics[width=0.40\textwidth]{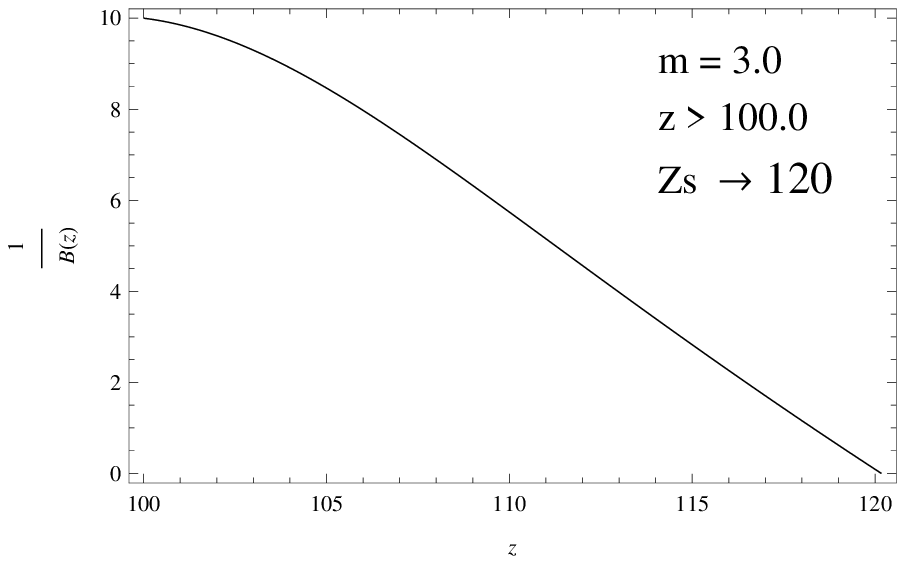}
\includegraphics[width=0.40\textwidth]{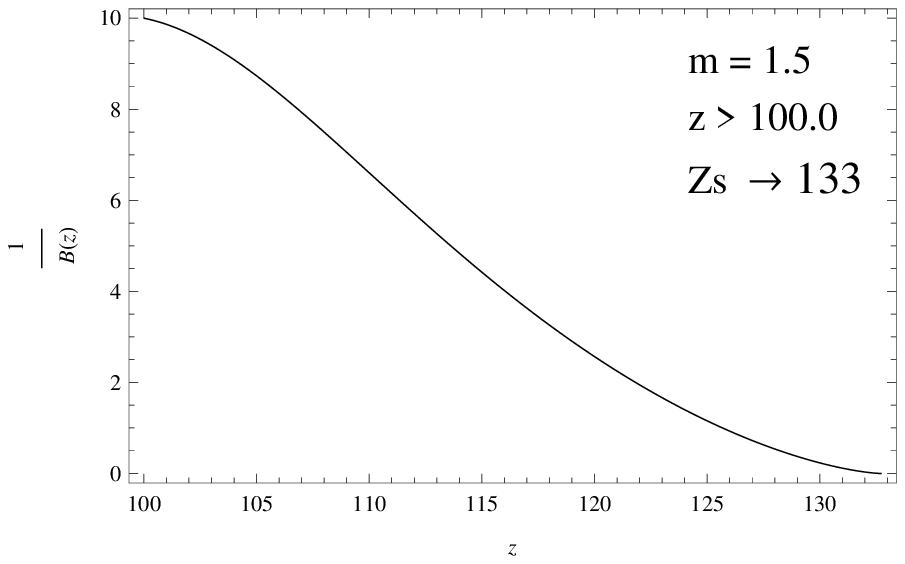}
\includegraphics[width=0.40\textwidth]{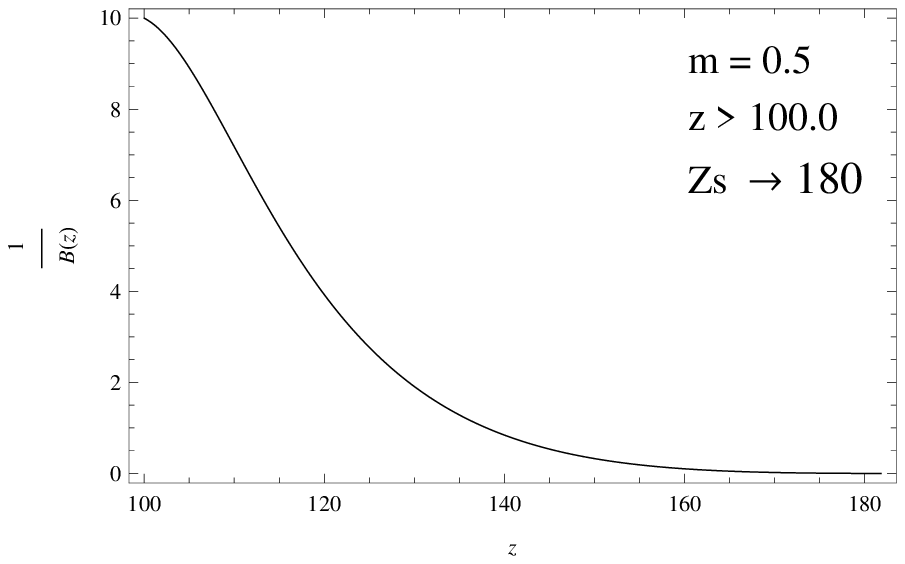}
\caption{Evolution of $\frac{1}{B(z)}$ with respect to $z$ for different choices of potential where $m > 0$: $m = 3.0, 1.5, 0.5$ i.e. $V(\phi) = \frac{\phi^2}{2} + 2\frac{\phi^{5/2}}{5}$, $V(\phi)= \frac{\phi^2}{2}+\frac{\phi^4}{4}$ and $V(\phi)= \frac{\phi^2}{2} + 2\frac{\phi^{3/2}}{3}$ respectively; for different initial conditions.}
\end{center}
\label{finfig}
\end{figure}

\begin{figure}[h]\label{RK42}
\begin{center}
\includegraphics[width=0.40\textwidth]{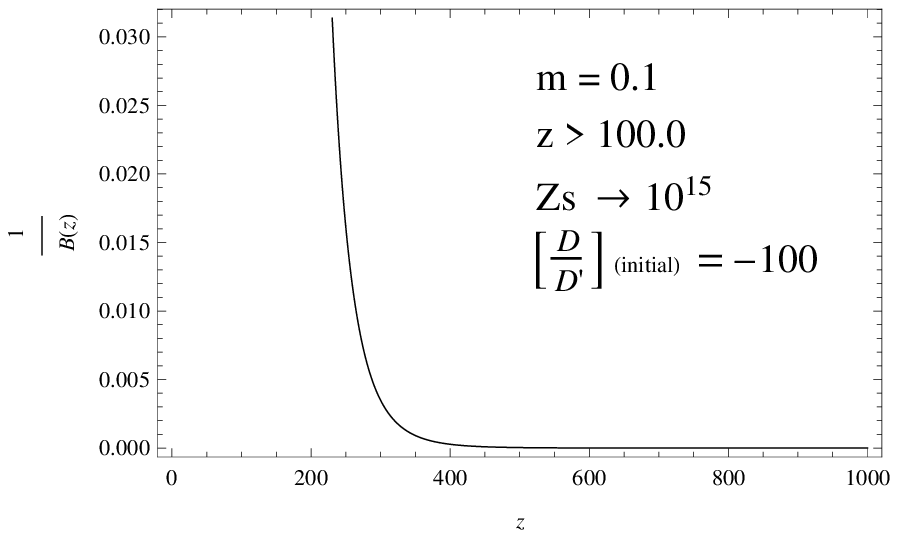}
\includegraphics[width=0.44\textwidth]{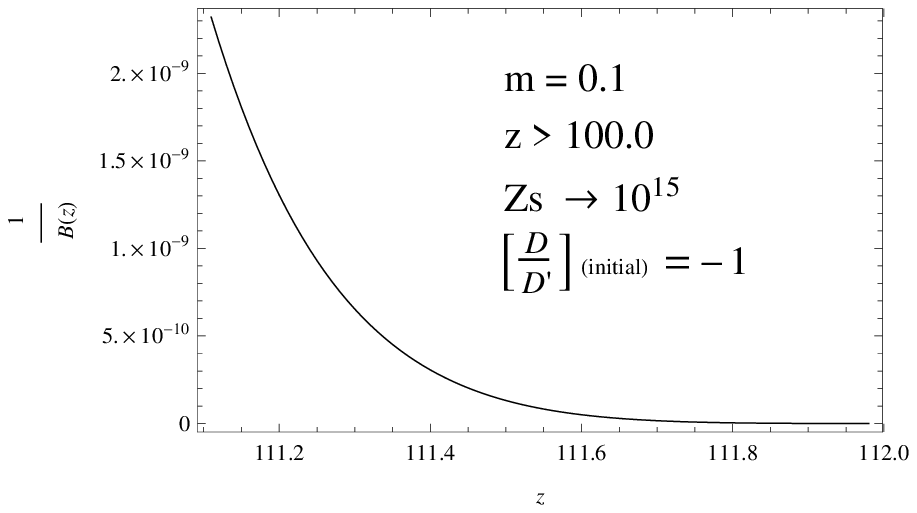}
\caption{Evolution of $\frac{1}{B(z)}$ with respect to $z$ for $m = \frac{1}{10}$ or $V(\phi)= \frac{\phi^2}{2} + 10\frac{\phi^{11/10}}{11}$. The case where $\mid D^{\circ}(z)\mid << D(z)$ is plotted on the LHS and the case where $\mid D^{\circ}(z)\mid \sim D(z)$ is on the RHS.}
\end{center}
\label{finfig}
\end{figure} 

\begin{enumerate}
\item{ For all $m > 0$, the spherical body collapses to a zero proper volume singularity at a finite future defined by $z = z_{s}$ as shown in figure $(5)$. For a large positive value of $m$, e.g $m \geq 3$, the singularity is reached more-or-less steadily. However, for choices of $m$ smaller in magnitude (for example, $m = 1.5, m = 0.5$), the system collapses at an increasingly greater value of $z_{s}$ and the evolution starts to look like an asymptotic curve. However, eventually the system attains a zero proper volume but at a very large but finite value of $z$, provided $m > 0$. This qualitative behaviour is independent of the initial condition defined by different choices of $\mid\frac{D(z)}{D^{\circ}(z)}\mid$.
}

\item{ When $m$ is a very small positive number (for example {\bf $m \sim \frac{1}{10}$}), the rate of collapse becomes increasingly sensitive to the initial condition $\mid\frac{D(z)}{D^{\circ}(z)}\mid$. The ultimate qualitative behaviour remains the same however; the system reaches a zero proper volume at a finite but very large $z_{s}$. For $\frac{D(z)}{D^{\circ}(z)} \sim -100$, the system falls very rapidly as suggested by figure $(6)$, followed by a stage when the rate of collapse is slowed down significantly and the singularity is not reached until $z \sim 10^{15}$. The characteristic difference with the case when $\frac{D(z)}{D^{\circ}(z)} \sim -1$ is clear from figure $(6)$.
}

\item{ For $-3 < m < 0 (m \neq -3,-1)$, the collapsing system approaches a zero proper volume with respect to $z$ asymptotically i.e. only when $z \rightarrow \infty$. In figure $(7)$, two examples are studied, for $m = -0.1$ and $m = -2.0$. The slope of the curves, i.e. the rate at which the spherical body approaches singularity may be different for different choices of $m$ in this domain, due to different signatures of the nonlinearities in equation (\ref{devolu2}). The collapse occurs more rapidly when $m = -0.1$ than $m = -2.0$. This nature is independent of the initial condition, i.e., whether $\mid D^{\circ}(z)\mid << D(z)$ or $\mid D^{\circ}(z)\mid \sim D(z)$.   
}

\item{ For $m < -3$, the evolution of the sphere is extremely sensitive to the choice of initial value of $D^{\circ}(z)$ with respect to $z$. An example is given in figure $(8)$, where we have studied the scenario for $m = -7.0$, or $V(\phi)= \frac{\phi^2}{2} - \frac{1}{6\phi^6}$ for different types of initial conditions. For some initial condition, when $\mid D^{\circ}(z)\mid \sim D(z)$, the sphere after an initial steady collapsing epoch, falls very sharply with respect to $z$; eventually hitting the zero proper volume singularity at a finite future. However, for $\mid D^{\circ}(z)\mid << D(z)$, the system can sometimes exhibit a somewhat oscillatory evolution as shown on the LHS of figure $(8)$. The oscillatory motion is followed by a very rapid, almost instanteneous drop to zero proper volume singularity.
}
\end{enumerate}

\begin{figure}[h]\label{RK43}
\begin{center}
\includegraphics[width=0.40\textwidth]{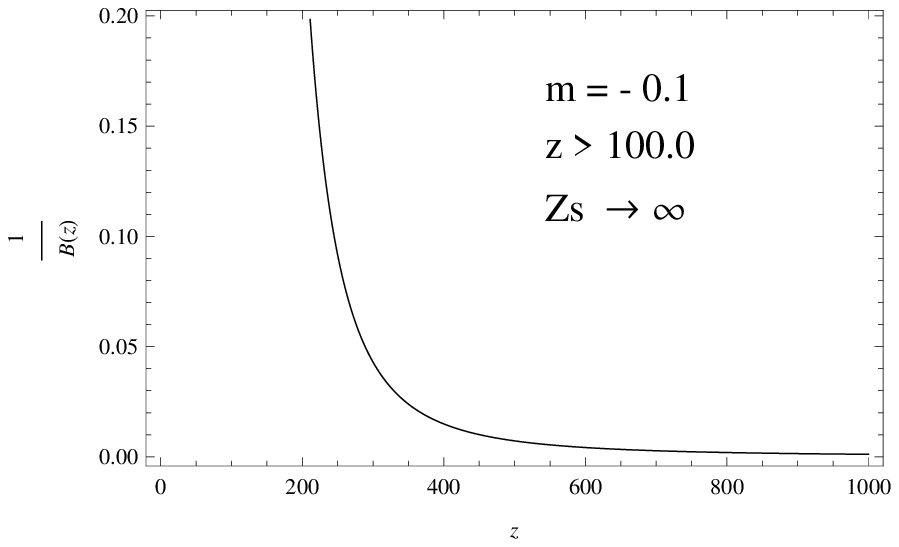}
\includegraphics[width=0.40\textwidth]{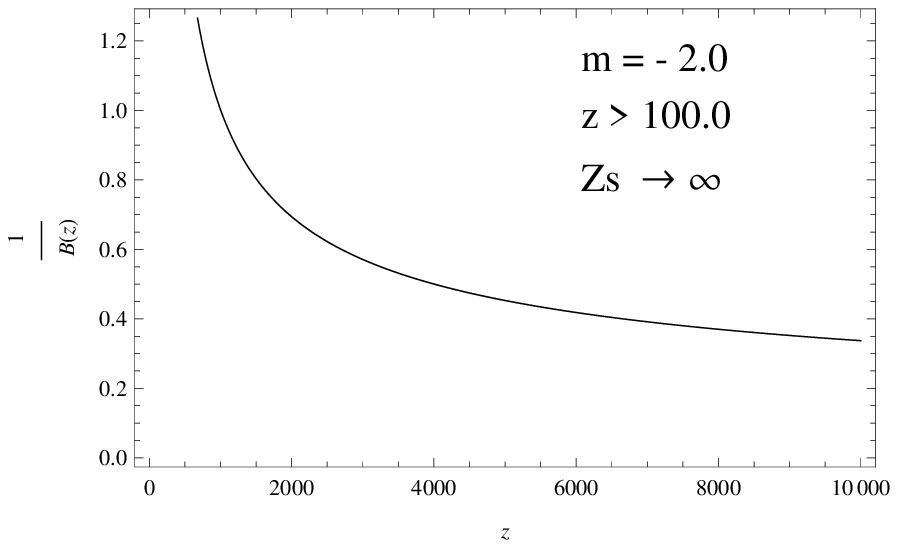}
\caption{Evolution of $\frac{1}{B(z)}$ with respect to $z$ for $m = -\frac{1}{10}$ or $V(\phi)= \frac{\phi^2}{2} + 10\frac{\phi^{9/10}}{9}$ and for $m = -2.0$ or $V(\phi)= \frac{\phi^2}{2} - \frac{1}{\phi}$ for different initial conditions defined by $\mid D^{\circ}(z)\mid << D(z)$ or $D^{\circ}(z) \sim D(z)$. The qualitative behaviour seems to be independent of the initial choice of parameters.}
\end{center}
\label{finfig}
\end{figure}

\begin{figure}[h]\label{RK44}
\begin{center}
\includegraphics[width=0.40\textwidth]{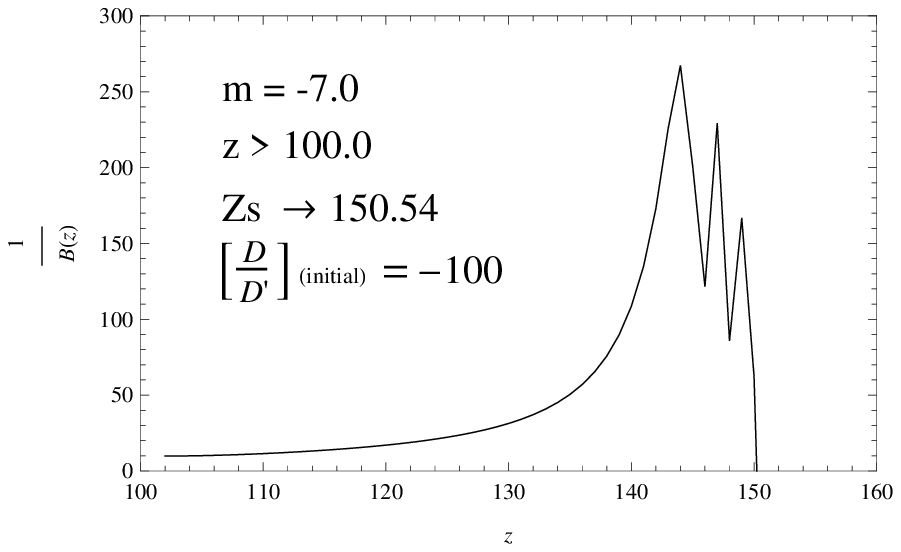}
\includegraphics[width=0.40\textwidth]{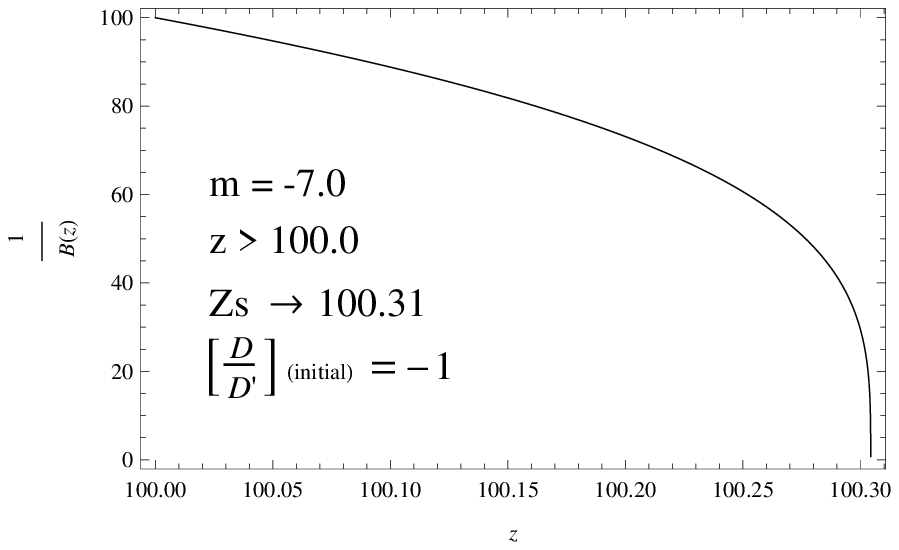}
\caption{Evolution of $B(z)$ with respect to $z$ for $m = -7.0$, or $V(\phi)= \frac{\phi^2}{2} - \frac{1}{6\phi^6}$ for different initial conditions.}
\end{center}
\label{finfig}
\end{figure}

\section{Discussion}
The collapse of a massive scalar field distribution is investigated in the present work. In a recent work\cite{scnb} a scheme for such an investigation is presented for a conformally flat spacetime. The basic method was to look for the integrability condition for the Klein-Gordon equation for the scalar field. The integrability condition for an anharmonic oscillator, as developed by Euler\cite{euler} is found to be quite apt to be used in this context. The present work includes a lot more generalization in the matter content in the sense that it has an anisotropic pressure and a heat flux as well. For this generalization regarding the matter field, another symmetry requirement, that of a self similarity, has been imposed so as to overcome the difficulty in integrating the scalar field equation. Without actually solving the set of Einstein equations, a lot of informations regarding the metric can be obtained where the scalar field equation is integrable.             \\

The scalar potential is assumed to be a power law one, which finds a lot of interest in cosmology, Except for a few powers, restricted by the domain of applicability of Eulers theorem, the method in fact applies for any other power law. We have picked up a few powers as examples, namely, ${\phi}^{4}$, ${\phi}^{\frac{2}{3}}$ and ${\phi}^{-\frac{1}{2}}$.      \\

Indeed, collapsing modes leading to a final singular state, are found with the help of numerical plots. Whether the singularity of a zero proper volume occurs at a finite future or the modes are for ever collapsing without practically hitting the singular state depends on the potential as well as the initial conditions. In order to facilitate further analysis like the formation of apparent horizon etc., we also do some analytical study with the help of some reasonable approximate solutions. The qualitative features of the analytical solutions are similar to the numerical solutions.        \\ 

Some of the singularities are found not covered by an apparent horizon. As already mentioned in section $5$, this apparent disagreement with the work of Hamid, Goswami and Maharaj\cite{hamid} perhaps lies in the fact that the scalar field contribution can violate the energy conditions. Anisotropy of the fluid pressure and the heat flux (departure from the perfect fluid) can also contribute towards this existence of naked singularities.

A ${\phi}^{2}$ potential is excluded from the purview of the investigation as this corresponds to $n=1$ which falls in the disallowed category as discussed in section $4$. We have included an investigation of a potential of the form $V = a {\phi}^{2} + b {\phi}^n$. In this case we could obtain only some numerical plots for the scale factor. For various choices of $n$, collapsing modes are found, and all of them reaches the singularity sooner or later. \\

It also deserves mention that the particular potentials chosen as examples have definite physical motivations as discussed in the introduction. But the condition for integrability stems from a mathematical interest, and no definite physics is associated with this. Still they represent physical situations, and a comprehensive study of scalar field collapse is possible.


\vskip 1.0cm

\begin{thebibliography}{99}
\bibitem{Opp} J. R. Oppenheimer and H. Snyder, Phys. Rev. {\bf 56}. 455 (1939).
\bibitem{pankaj1} P. S. Joshi, {\it Global aspects in Gravitation and Cosmology}; Clarendon Press, Oxford (1993).
\bibitem{pankaj2} P. S. Joshi, arXiv:1305.1005.
\bibitem{paddy} T. Padmanabhan, Phys. Rep. {\bf 380}, 235 (2003).
\bibitem{varun} V. Sahni and A. Starobinsky, Int. J. Mod.Phys. D {\bf 9}, 373 (2000).
\bibitem{sami} E. J. Copeland, M. Sami and S. Tsujikawa, Int. J. Mod. Phys. D {\bf 15}, 1753 (2006).
\bibitem{mota} D. F. Mota and C. Van de Bruck, Astron. Astrophys. {\bf 421}, 71 (2004).
\bibitem{gong2} S. Goncalves and I.Moss, Class. Quant. Grav.; {\bf 14}, 2607 (1997).
\bibitem{christo1} D. Christodoulou, Commun. Math. Phys., {\bf 109}, 591 (1987); {\bf 109}, 613 (1987).
\bibitem{christo3} D. Christodoulou, Ann. Math., {\bf 140}, 607 (1994).
\bibitem{piran} D. Goldwirth and T. Piran, Phys. Rev. D; {\bf 36}, 3575 (1987).
\bibitem{goswami} R. Goswami and P. S. Joshi, Phys. Rev. D; {\bf 69}, 027502 (2004).
\bibitem{giambo} R. Giambo, Class. Quant. Grav; {\bf 22}, 2295 (2005).
\bibitem{gong1} S. Goncalves, Phys. Rev. D; {\bf 62}, 124006 (2000).
\bibitem{goswami2} R. Goswami and P. S. Joshi, Mod. Phys. Lett. A; {\bf 22}, 65(2007).
\bibitem{koyel} K. Ganguly and N. Banerjee, Pramana, {\bf 80}, 439, (2013).
\bibitem{cai1} R.-G. Cai and R.-Q. Yang, arXiv:1602.00112.
\bibitem{cai2} R.-G. Cai, L.-W. Ji and R.-Q. Yang, arXiv:1612.07095.
\bibitem{chop} M. W. Choptuik, Phys. Rev. Lett.; {\bf 70}, 9(1993).
\bibitem{brady} P. R. Brady, Class.Quant.Grav.; {\bf 11}, 1255 (1994).
\bibitem{gund} C. Gundlach, Phys. Rev. Lett.; {\bf 75}, 3214(1995).
\bibitem{Gundlach1} C. Gundlach, “Critical phenomena in Gravitational Collapse: Living Reviews”; LivingRev. Rel. 2:4(1999).
\bibitem{samibook} M. Sami, Lecture Notes in Physics; "The Invisible Universe: Dark Matter and Dark Energy": {\bf 720}, 219(2007).
\bibitem{zlatev} P.J. Steinhardt, L.-M. Wang and I. Zlatev, Phys. Tev. D, {\bf 59}, 123504 (1999). 
\bibitem{johri} V. B. Johri, Phys. Rev. D, {\bf 63}, 103504 (2001).
\bibitem{scnb} S. Chakrabarti and N. Banerjee, arXiv:1609.01868 [gr-qc].
\bibitem{euler} N. Euler, Journal of Nonlinear Mathematical Physics {\bf 4}, 310 (1997).
\bibitem{harko} T. Harko, F. S. N. Lobo, M. K. Mak, Journal of Pure and Applied Mathematics: Advances and Applications {\bf 10 (1)} (2013) 115.
\bibitem{carr} B.J. Carr and A.A. Coley, Class. Quantum Grav., {\bf 16}, R31 (1999).
\bibitem{som} M. M. Som and N. O. Santos, Phys. Lett. A {\bf 87}, 89 (1981).
\bibitem{maiti} S. R. Maiti, Phys. Rev. D {\bf 25}, 2518 (1982).
\bibitem{modak} B. Modak, J. Astrophys. Astron. {\bf 5}, 317 (1984).
\bibitem{bhui} A. Banerjee, S. Choudhury and B. Bhui, Phy. Rev. D {\bf 40}, 670 (1989).
\bibitem{patel} L. K. Patel and R. Tikekar, Mathematics Today {\bf IX}, 19 (1991).
\bibitem{schafer} D. Schafer and H. F. Goenner, Gen. Relativ. Grav. {\bf 32}, 2119 (2000).
\bibitem{ivanov} B. V. Ivanov, Gen. Relativ. Grav. {\bf 44}, 1835 (2012).
\bibitem{herrera} L. Herrera, G. Le Denmat, N. O. Santos and A. Wang, Int. J. Mod. Phys. D {\bf 13}, 583 (2004).
\bibitem{santos1}L. Herrera, N. O. Santos, Physics Reports, {\bf 286}, 53, (1997).
\bibitem{herre2} L. Herrera, A. Di Prisco, J. Martin, J. Ospino, N.O.Santos, O.Troconis, Phys. Rev. D {\bf 69} : 084026, {\bf 2004}.
\bibitem{herre3} L. Herrera and J. Ponce de Leon , J. Math. Phys. {\bf 26}, 2018 (1985).
\bibitem{kaza} D. Kazanas and D. Schramm, Sources of Gravitational Radiation, L. Smarr ed., (Cambridge University Press, Cambridge, 1979).
\bibitem{maartens} R. Maartens and S.D. Maharaj, Class. Quantum Grav., {\bf 3}, 1005 (1986).
\bibitem{dwivedi} P. S. Joshi, I. H. Dwivedi: Class. Quant. Grav. \textbf{16} (1999) 41.
\bibitem{santos} N. O. Santos, Mon. Not. R. Astron. Soc. {\bf 216}, 403 (1985).
\bibitem{chan} R. Chan, Mon. Not. R. Astron. Soc. {\bf 316}, 588 (2000).
\bibitem{maharaj} S. D. Maharaj, M. Govender, Int. J. Mod. Phys. D {\bf 14}, 667 (2005).
\bibitem{ori} A. Ori, T. Piran, Phys. Rev. Lett. {\bf 59}, 2137, (1987).
\bibitem{waugh} B. Waugh, K. Lake, Phys. Rev. D. {\bf 38}, 1315, (1988).
\bibitem{waugh2} B. Waugh, K. Lake, Phys. Rev. D. {\bf 40}, 2137, (1989).
\bibitem{jdw1} P. S. Joshi, I. H. Dwivedi, Commun. Math. Phys. {\bf 146}, 333 (1992).
\bibitem{jdw2} I. H. Dwivedi, P. S. Joshi, Commun. Math. Phys. {\bf 166}, 117 (1994).
\bibitem{dixit} I. H. Dwivedi, S. Dixit, Prog. Theor. Phys. {\bf 85}, 433, (1991).
\bibitem{goswami3} R. Goswami, P. S. Joshi,  Phys. Rev. D {\bf 76}, 084026, (2007).
\bibitem{ghosh} S. G. Ghosh, D. W. Deshkar, International Journal of Modern Physics D.
{\bf 12}, 5 (2003) 913.
\bibitem{dadhi} P. S. Joshi, R. Goswami and N. Dadhich, Phys. Rev. D, {\bf 70}, 087502 (2004).
\bibitem{hamid} A. I. M. Hamid, R. Goswami, S. D. Maharaj, Class. Quant. Grav. {\bf 31} (2014) 135010.
\bibitem{herre5} L. Herrera, A. Di Prisco, J. Ospino, Gen. Relativ. Gravit. {\bf 42}, 1585,(2010).

\end{thebibliography}
\end{document}